\documentclass[
    ,final            % use final for the camera ready runs
  ] {aipproc}

\layoutstyle{6x9}
\usepackage{amssymb}
\begin{document}

\title{Philosophical problems of space-time theories}


\author{Gustavo E. Romero}{
  address={Instituto Argentino de Radioastronom{\'{i}}a (IAR, CCT La Plata, CONICET), C.C. No. 5, 1894, Villa Elisa, Buenos Aires, Argentina.\\ Email: romero@iar-conicet.gov.ar }
}

\begin{abstract}
 I present a discussion of some open issues in the philosophy of space-time theories. Emphasis is put on the ontological nature of space and time, the relation between determinism and predictability, the origin of irreversible processes in an expanding Universe, and the compatibility of relativity and quantum mechanics. In particular, I argue for a  Parmenidean view of time and change, I make clear the difference between ontological determinism and predictability, propose that the origin of the asymmetry observed in physical processes is related to the existence of cosmological horizons, and present a non-local concept of causality that can accommodate both special relativity and quantum entanglement.   
\end{abstract}

\maketitle

%%%%%%%%%%%%%%%%%%%%%%%%%%%%%%%%%%%%%%%%%%%%
%% MAINMATTER
%%%%%%%%%%%%%%%%%%%%%%%%%%%%%%%%%%%%%%%%%%%%

\section{Introduction}
\subsection{ Why physicists should care about philosophy?}

Many physicists do not  have a high opinion of the philosopher's work or about what philosophy can contribute to clarify the foundations of physical theories. This attitude is probably the result of some loose ways of doing philosophy since the 19th  century. This discomfort towards some philosophical speculations was even shared  by some philosophers. Let us see, for instance, what Schopenhauer has to say about Hegel:

 \begin{quotation}
Hegel was commonplace, inane, loathsome, repulsive and an ignorant charlatan, who with unparalleled effrontery compiled a system of crazy nonsense that was trumpeted abroad as immortal wisdom by his mercenary followers...
\end{quotation}

If philosophers have such thoughts about each other, why should we mind what they say about physics? Actually, the distinction between physics and philosophy is rather recent. Until the 19th century `natural philosophy' was the usual term to describe the research and study of nature. Science was initiated in Ancient Greece by people such as Thales, Anaximander, and other pre-Socratic thinkers who speculated about the nature of the physical world and were ready to criticize and argue rationally about their conjectures. The founders of modern science, like Nicolas Oresme, Galileo Galilei, and Isaac Newton were prone to philosophical and theological speculations. They were also worried about the very foundations of the theories they proposed and their interpretations. Many thinkers of the 17th century, including Descartes, Pascal, Huygens, and Leibniz can be considered either as scientists or philosophers, attending different aspects of their investigations and concerns. Even in the 18th century, Kant,  who was probably the most influential philosopher of his time, was well-learned in physical science and made some contributions to it. 

It is only with the advent of professional philosophy that a divergence started, in the early 19th century. Some philosophers, influenced by the Romantic movement, cultivated a disdain for science and the clear expression of thought. Approximately at the same time, however, a revolution in mathematics lead by Gauss, Riemann, Weierstrass, and others motivated many mathematicians to become interested in in philosophical problems following a tradition that can be traced to Pythagoras, Parmenides, and Zeno of Elea. William K. Clifford, John Stuart Mill, Friedrich L. Gottlob Frege, Charles S. Peirce, and many others are clear examples of the quest for clarity of expression and rational thinking in the 19th century. All this movement led to the birth of  mathematical logic and formal languages especially suited for the philosophical discussion of modern science; a formidable task that was systematically pursued in the 20th century by thinkers such as Bertrand Russell, David Hilbert, Henri Poincar\'e, Hermann Weyl, Moritz Schlick, Rudolph Carnap, Hans Reichenbach, and many others. 

Logical empiricism gave rise to an unprecedented standard of precision and formalization in the philosophical consideration of scientific achievements and theories. Under the general expression of `analytic philosophy', a scientifically informed way of research in philosophy spread especially in the English-speaking world after the Second World War. Currently, the logical and semantic tools at the disposal of analytical philosophers seem indispensable in solving the interpretation problems of several theories of contemporary physics. 

Some scientists soon undertook this, and many of the greatest physicists of the 20th century made active research on philosophical and interpretation topics of quantum mechanics and general relativity. Among them we can highlight Einstein, Bohr, Heisenberg, Schr\"odinger, Hoyle, and Penrose; all of them wrote extensively on different philosophical problems.     

All physical theories assume some ontology (i.e. some entities that are considered as existent in the actual world), and some formal structure that requires proper interpretation. So, consciously or not, all physicists make philosophical commitments when doing scientific research. It is better if these assumptions are explicit and consistent. This allows a better understanding of the inner structure of scientific theories and facilitates their confrontation with facts. A clear and consistent interpretation, in addition, is fundamental to prompt new developments. Contrarily, inconsistent or ad-hoc interpretations can hinder the progress of science (see Ref. \cite{Bunge1967} for a full discussion).  

Perhaps there are no more basic concepts in physics than those of space and time. All physical theories presuppose them, even the so-called `space-time' theories, like General Relativity. Space-time theories deal with the structure and representation of space and time, but say nothing about their nature. Are space and time physical entities, like electrons or stars? Are they mere forms of our perception of such objects, as Kant thought? Do they exist at all? If they exist, what are they made of? These are the kind of questions that we will discuss first in our brief survey of philosophical topics of modern physics.  

\section{The nature of space and time}

\subsection{A brief history of the problem}

Concern for the nature of space and time can be traced to Ancient Greece. Natural philosophers in the 5th century BC were worried about the origin of change. For Heraclitus of Ephesus, change was a basic feature of nature: everything was `in flux'. Even things that apparently are immutable, at a basic level, according to Heraclitus, are undergoing change. Parmenides of Elea famously denied the mere possibility of change. He pointed out that `what is, is' and `what is not, is not'. `Nothing' is \emph{not} a thing. It is just the denial of the existence of things and, then, it cannot be an existent itself.  Hence, void, or empty space, cannot exist. Without empty space, for Parmenides, motion was impossible. He understood motion as the successive occupation of empty places. If everything is filled, there is nowhere to move on.  Reality must be a single, homogeneous, unchanging unity. If we think that there is change, it is just because we do not have a true perception of reality. Plato was deeply influenced by Parmenides, and postulated a world of unchanging ideas, and a world of shadows where the physical, changing objects, were just a mere reflection of the true world. Change, for Plato, is a form of corruption. Time, the mobile form of eternity. Ontologically, for Plato, eternity is prior to time \cite{Borges}. 

The challenge of Parmenides was taken at face value by Democritus and Leucippus of Abdera. They accepted Parmenides's idea that there cannot be change if there is no void, and they inverted the reasoning: since there is change, then there must be a void. Democritus and Leucippus, therefore, postulated two kind of existents: atoms and the void. Atoms are indivisible and do not change. Void separates atoms so they can move and combine to form the things that populate the Universe. Actually, since there are an infinite number of atoms, there should be an infinite number of worlds. Ours is just one among them. Change, then, is an emerging property of complex things. Time, a measure of change. Interestingly, time results as an emergent property as well. If there were no changing things, there would be no time. 

Aristotle, in the 4th century BC, adopted a different theory of change, but kept the same conclusion: time is the measure of change; space, the measure of extension. Nothing would exist without changing things. For Aristotle, however, there are no atoms, but substances and forms. Any object is some substance (the `stuff' that makes it) with some form. Substance remains, but form can change.   

 The logical treaties of Aristotle were preserved along the early Middle Ages, but the physical and metaphysical treaties were recovered only in the High Middle Ages (12th-13th centuries). Platonic views, which were dominant so far, slowly retreated in front of the natural philosophy of Aristotle and his commentators. Different aspects of his physics were developed and criticized by scholars like John Buridam, Nicolas Oresme, William of Ockham, and Roger Bacon \cite{Lindberg}. By the early 17th century, after the work of  Kepler, Copernicus, and Galileo, the Western natural philosophy was on the verge of a major revolution. 

This revolution, introduced by Isaac Newton, would be the culmination of a process of understanding the world initiated in the Middle Ages with the revision of Aristotelian views. It was a rediscovery of the theoretical, hypothetical-deductive methods used by Parmenides, now powered by new mathematical tools. It was already clear for Oresme and other scholars that kinematic and dynamics were quite different aspects of change. Newton was able to develop a quantitative theory of dynamics and a theory of gravitation. He could introduce constraints to the motion of physical objects with differential equations for the functions that represent the position of the bodies. Time appears as a free parameter in these equations. In his \emph{Philosopiae Naturalis Priincipia Mathematica} he famously wrote:

\begin{quotation}
Absolute, true, and mathematical time, from its own nature, passes equably without relation to anything external, and thus without reference to any change or way of measuring of time. 
\end{quotation}

  \begin{quotation}
Absolute, true, and mathematical space remains similar and immovable without relation to anything external. 
\end{quotation}

\begin{quotation}
The place of a body is the space which it occupies, and may be absolute or relative according to whether the space is absolute or relative.
\end{quotation}

\begin{quotation}
Absolute motion is the translation of a body from one absolute place to another; relative motion the translation from one relative place to another.
\end{quotation}

It is evident from these characterizations that, according to Newton: 
\begin{itemize}
\item Space is something distinct from a body and exists independently of the existence of bodies.
\item Time is a kind fluid, that exists independently of any body.
\item The true motion of a body does not consist of, or cannot be defined in terms of, its motion relative to other bodies.
\end{itemize}

Actually, absolute motion for Newton is motion relative to absolute space . The origin of his ideas was perhaps in a work by William Charleton which appeared in 1654 and was entitled \emph{ Physiologia Epicuro-Gassendo-Charltoniana: Or a Fabrick of Science Natural, upon the Hypothesis of Atoms, ``Founded by Epicurus, Repaired by Petrus Gassendus, Augmented by Walter Charleton"}. In that work, Charleton revisited the Greek atomic doctrine and stated that  time and space are real entities even though they fit neither of the traditional categories of substance nor accident (i.e., property of a substance). Additionally, he claimed that  time ``flow[s] on eternally in the same calm and equal tenor," while the motion of all bodies is subject to "acceleration, retardation, or suspension".

 The postulation of two entities, space and time, without any obvious measurable properties, was suspicious for several thinkers. Gottfried W. Leibniz, in his correspondence with the British thinker Samuel Clarke \cite{Ariew}, defended a relationalist view of space and time and attacked the absolute conceptions of Clarke and Newton.  Leibniz was a major continental rationalist. He  used logic rather than empirical data to attack Newton's views on space and time. His arguments were based in two principles, that he considered self-evident: the Principle of Sufficient Reason (PSR) and the Principle of the Identity of Indiscernibles (PII). In his own words:
\begin{itemize}

\item PSR: ``there ought to be some sufficient reason why things should be so, and not otherwise''.

\item PII: ``to suppose two things indiscernible, is to suppose the same thing under two names''.

\end{itemize}

From these principles, he developed the following argument against absolute space. Let us consider two universes with exactly the same things in them located in the exact same spatial relations, but, imagine, that these two set of things are located, altogether, in different absolute locations in the two universes. Now, the two universes are indiscernible. So, according to PII they should be the same. But they are not because we are assuming an absolute space. The two universes also contradict the PSR, since why should one of the universes exist instead of the other? These contradictions are solved just dropping the concept of an absolute space. 

Instead of absolute space and time, the notions which he considered unintelligible, Leibniz proposed relational concepts:   ``I hold space to be something merely relative, as time is, that I hold it to be an order of coexistences, as time is an order of successions''. Space, is then understood as a set of relations among things. Actually, there is no space in itself, there are just spatial relations among things. Empty space is an impossible thing, a contradiction. For this reason, for Leibniz, the Universe is a \emph{plenum}, as it was for Parmenides. However, Leibniz not only accepts change. For him, change is essential. Time is just a  name for relative change. If there were no changing things, there would be no time. 

Newton argued for absolute space and motion through a thought experiment. Imagine a rotating bucket filled with water. After a span communicating the motion of the bucket, the water is again at rest with respect to it, but now its surface is concave. In this way, absolute circular motion (acceleration)  can be detected. Leibniz died in 1716, without providing a satisfactory response to this remark by Newton, but with the conviction that ``the immediate cause of the change is in the body itself''.

In the 19th century Ernst Mach also had the opinion that absolute motion makes non-sense. In 1893 he wrote \cite{Mach}:

\begin{quotation}
Newton's experiment with the rotating water bucket teaches us only that the rotation of water relative to the bucket walls does not stir any noticeable centrifugal forces; these are prompted, however, by its rotation relative to the mass of the Earth and the other celestial bodies. Nobody can say how the experiment would turn out, both quantitatively and qualitatively, if the bucket walls became increasingly thicker and more massive and eventually several miles thick.
\end{quotation}  

For Mach, the influence of the `distant stars', i.e. the rest of the mass in the Universe, is responsible for the curvature of the water's surface. Were the entire Universe rotating around the bucket, the effect would be the same. All motion is \emph{always} relative. 

In the early 20th century, these ideas would lead Poincar\'e and then Einstein to formulate the Special Theory of Relativity. The theory is based on two simple postulates: 1. The speed of light in vacuum is absolute, and 2. The laws of physics are same in all systems in relative constant motion. The consequence is that kinematics and dynamics are not any more Galilean-invariant, but, as electromagnetic equations, Lorentz-invariant. This leads to a dramatic reformulation of the concept of simultaneity, that now depends on the inertial reference  frame \cite{Einstein1}, and then to the introduction of space-time in physics.

\subsection{Space-time}

On September 21st 1908, Hermann Minkowski addressed the audience of the  80th Assembly of German Natural Scientists and Physicians in Cologne with these rightly famous opening words \cite{Minkowski}:

\begin{quotation}
The views of space and time which I wish to lay before you have sprung from the soil of experimental physics, and therein lies their strength. They are radical. Henceforth space by itself, and time by itself, are doomed to fade away into mere shadows, and only a kind of union of the two will preserve an independent reality.
\end{quotation}  

What is this space-time of Minkowski?\\

{\sl Space-time is the ontological sum of all events of all things}. \\

A thing is an individual endowed with physical properties. An event is a change in the properties of a thing. An ontological sum is an aggregation of things or physical properties, i.e. a physical entity or an emergent property. An ontological sum should not be confused with a set, which is a mathematical construct and has only mathematical (i.e. fictional) properties. 

Everything that has happened, everything that happens, everything that will happen, is just an element, a ``point'', of space-time. Space-time is not a thing, it is just the relational property of all things.

As happens with every physical property, we can represent space-time with some mathematical structure, in order to describe it. We shall adopt the following mathematical structure for space-time:\\

{\sl Space-time can be represented by a $C^{\infty}$ differentiable, 4-dimensional, real manifold.}\\

A real 4-D manifold is a set that can be covered completely by subsets whose elements are in a one-to-one correspondence with subsets of $\Re^{4}$. Each element of the manifold represents an event. We adopt 4 dimensions because it seems enough to give 4 real numbers to localize an event. We can always provide a set of 4 real numbers for every event, and this can be done independently of the intrinsic geometry of the manifold. If there is more than a single characterization of an event, we can always find a transformation law between the different coordinate systems. This is a basic property of manifolds. 

Now, if we want to calculate distances between two events, we need more structure on the manifold: we need a geometric structure. We can get it introducing a metric tensor $g_{\mu\nu}$ to determine distances. The infinitesimal separation between two events of space-time is given by:

\begin{equation}
ds^{2}=g_{\mu\nu} dx^{\mu} dx^{\nu}.
\end{equation}

Space-time, then, is fully represented by an order pair $(M, g)$, where $M$ is the manifold and $g$ is the metric tensor. In General Relativity \cite{Einstein2}, the metric of space-time is determined by the energy-momentum of the physical systems. These properties are represented by the so-called energy-momentum tensor. The metric itself then represents the gravitational potential and its derivatives determine the equations of motion through the affine connection of the manifold. 

If Leibniz was right in his views of space and time, then it should be possible to build space-time and its metrical properties from the properties of things. Space-time would naturally emerge as a global relational property of the system of all things. If things have quantum properties then space-time should reflect this at certain level. For discussions and steps in this direction see Refs. \cite{Rovelli} and \cite{Bergliaffa}.

\subsection{Can we know the true geometry of the world?}

In the late 19th century Henri Poincar\'e  held that convention plays an important role in physics. His view (and some later, more extreme versions of it) came to be known as ``conventionalism''.  Poincar\'e believed that the geometry of physical space is conventional. He considered examples in which either the geometry of the physical fields or gradients of physical properties can be changed, either describing a space as non-Euclidean measured by rigid rulers, or as a Euclidean space where the rulers are expanded or shrunk by a variable heat distribution. Poincar\'e, however, thought that as we were so used to Euclidean geometry, we would prefer to change the physical laws to save Euclidean geometry rather than shift to a non-Euclidean physical geometry. This conventionalist approach was applied also to the interpretation of Special Relativity, and later to General Relativity (e.g. by Reichenbach and Gr\"unbaum \cite{Reichenbach}, \cite{Grunbaum}).

Reichenbach argued, in the vein of Poincar\'e, that an arbitrary geometry may be ascribed to space-time (holding constant the underlying topology) if the laws of physics are correspondingly modified through the introduction of ``universal forces''. Reichenbach's thesis of metrical conventionalism is part of a  program of epistemological reductionism regarding space-time structures, where metrical properties of space-time are deemed less fundamental than topological ones. The end point of Reichenbach's epistemological analysis of the foundations of space-time theory is ``the causal theory of time'', a type of relational theory of time that assumes the validity of the causal principle of action-by-contact.

In response to Poincar\'e and the Reichenbach challenge we can argue that theories cannot be arbitrary complex in order to save the Euclidean geometry. Besides, the web of our theories about the world must be coherent. To adopt complex ad-hoc fields to compensate non-Euclideaness can have undesirable effects in other physical theories, like quantum mechanics. In brief, we can say that all entities postulated by a theory are just conjectural and can be adjusted when the consequences of the theory are confronted with experiment. Our knowledge of the ontology of the world is not conventional but conjectural.

\subsection{A modest vindication of Paremenides: a block Universe}

Parmenides of Elea, a town on the west coast of southern Italy, lived from the end of the 6th century to the mid 5th century BC. He wrote a poem in the hexameter meter entitled {\em On What Is}. This piece contains the first known example of a deductive system applied to the physical reality. Parmenides was not content just with giving his view of the world. He supported his view by logical deduction from what he considered self-evident premises. He stated, as I have mentioned, that there is no change, no becoming, no coming to be. Reality turns out to be unchanging, eternal, motionless, perfect, and single. There is just one thing: the World. His monism is absolute. What we think is a changing world is only the result of illusion and deception. 

The premises of Parmenides's argument can be written as:

\begin{itemize}
\item What is, is.
\item What is not, is not.
\end{itemize}

Then, nothing can come to be from what is not, because `what is not' is not something. Change is impossible, since for Parmenides change is the occupation of empty space, but there cannot be `empty space'. Reality must be an unchanging block. 

The atomists Democritus and Leucippus denied the conclusion: there is no change. Hence, they inferred, one of the premises is wrong. They decided to reject the second one, and then stated that there are just two kind of things in the world: atoms and the void. They invented the first theory of change. 

Many centuries later, with the advent of field theories it became clear that change can occur even in a full Universe: change does not require empty space. A perturbation in a field that fills the whole Universe is a change. A change in degree or strength does not require empty space. 

The concept of change, as an ordered pair of states or an event, is central to the manifold model of space-time. But once the geometry of the manifold is given by a distribution of energy and momentum, its structure is fixed. The `points' of the manifold represent events, but there is no event or change affecting the space-time as a whole. The four dimensional Universe, represented by the manifold model  is unchanging, eternal, motionless, single, just as the Parmenidean Universe. What we call irreversible processes are asymmetries in the space-time. The objects that populate the Universe are 4-dimensional. They have `temporal parts', as well as spatial parts. In this way, the child I was, is just a part of a larger being, I, that is 4-dimensional. What we call `birth' and `death' are just temporal boundaries of such a being. Change appears only when we consider 3-dimensional slices of 4-dimensional objects. In words of Max Tegmark:

\begin{quotation}
Time is the fourth dimension. The passage of time is an illusion. We have this illusion of a changing, three-dimensional world, even though nothing changes in the four dimensional union of space and time of Einstein's relativity theory. If life were a movie, physical reality would be the entire DVD: Future and past frames exist just as much as the present one.
\end{quotation}

Therefore, it seems fair to call this interpretation of space-time, usually known as `the block Universe' \cite{Smart}, a Parmenidean view of the world. Parmenides is back with a vengeance, in 4 dimensions.

\subsection{When is `now'?}

Events in space-time do not flow. They simply {\em are}. What is called the transient `now' or `present' is not itself an event in space-time. Events are ordered by the relations `earlier than' or `later than', but no event is singled out as `present', except by convention. As noted by Adolf G\"unbaum \cite{Grunbaum}, what is `now' about a given event is that it is affecting some conscious being who is aware of the event. The `present' is not an intrinsic property of a given event, much less a changing thing in the world, but a relation among some number of events and a self-conscious individual.  More specifically, we can define: \\

{\bf Present (psychological)}: {\em class of all events that are physically related to given brain event.}\\
          
The present, consequently, like smell, sweetness, and other secondary qualities, is introduced by the interaction of sentient individuals with their environment. Time does not `flow' in any physical sense. What changes is the state of our consciousness of the events in our surroundings. Hermann Weyl (1885-1955) wrote:

\begin{quotation}
	The objective world simply is, it does not happen. Only the gaze of my consciousness, crawling upward along the life line of my body, does a section of this world come to life as a fleeting image in space which continuously  changes in time \cite{Weyl}. 
\end{quotation}

The awareness of present by a certain individual requires a series of time-like events in the brain, and a series of such events implies some time interval. The integration of all physically related events gives some temporal `thickness' to what we perceive as the `present'. This is what William James called the `specious present'.\\

{\bf Specious present}: {\em length of the time-history of brain processes necessary to integrate all events that are physically related to given brain event.}\\

The specious present, then, being related to a brain process, can be different for different individuals equipped with different brains.     

The `becoming' is not a property of the events of space-time, but a property of our consciousness of such events. We call `becoming' to the series of states of consciousness associated with a series of physical events. Summing up, time does not go by. We do.

%%%%%%%%%%%%%%%%%%%%%%%%%%%%%%%%%%%%%%%%%%%%
%% Sample figure:
%%
%% The option [height=...] scales the picture to the given height,
%% without it it would be printed at its nominal size
%%%%%%%%%%%%%%%%%%%%%%%%%%%%%%%%%%%%%%%%%%%%

%\begin{figure}
 % \includegraphics[height=.3\textheight]{golfer}
 % \caption{Picture to fixed height}
%\end{figure}

\section{Determinism and predictability}

Determinism is a metaphysical doctrine about the nature of the world. It is an ontological assumption: the assumption that all events are given. It can be traced to Parmenides and its ``what is, is''. It is important to emphasize that determinism does not require causality and does not imply predictability. 
Predictability is a property of our theories about the world, not a property of the world itself. 

The confusion between determinism and predictability can be traced to Pierre-Simon Laplace and his {\em Philosophical Essay on Probabilities}:

\begin{quotation}
We may regard the present state of the Universe as the effect of its past and the cause of its future. An intellect which at a certain moment would know all forces that set nature in motion, and all positions of all items of which nature is composed, if this intellect were also vast enough to submit these data to analysis, it would embrace in a single formula the movements of the greatest bodies of the Universe and those of the tiniest atom; for such an intellect nothing would be uncertain and the future just like the past would be present before its eyes.
\end{quotation}

According to Laplace, every state of the Universe is determined by a set of initial conditions and the laws of physics. Since the laws are represented by differential equations and there are theorems for the existence and uniqueness of solutions, determinism implies predictability. Theorems apply, however, only to mathematical objects, not to reality. The world is not mathematical, just some of our representations of it are mathematical. The existence of solutions to some equations that represent physical laws does not imply physical existence. Physical existence is independent of our conceptions. Moreover, even in Newtonian space-times there are Cauchy horizons. These are hypersurfaces from where, even the in case of a complete specification of initial data, the solutions of dynamical equations cannot predict all future events. This arises because of the absence of an upper bound on the velocities of moving objects. For instance, consider the trajectory of an object that is accelerated in such a way that its velocity becomes in effect infinite in a finite time. This object will be disconnected from all later times (see Ref. \cite{Earman} for more examples). 

General Relativity assumes the existence of all events represented by a manifold. Hence, it is an deterministic theory. 
The Cauchy problem, however, cannot always be solved in General Relativity. Cauchy horizons naturally appear in many solutions of Einstein field equations (from wormholes to black hole interiors). Although the manifold is fixed, we cannot always describe it from limited knowledge. A physical theory can be ontologically deterministic but nonetheless epistemologically underdetermined.  

The fact that there exist irreversible processes in the universe, implies that the space-time manifold is intrinsically asymmetric. The laws that constrain the space-state of physical things, and therefore their potential to change, however, are invariant under time reversal. We turn now to this problem.

\section{Irreversibility and space-time}

The Second Law of Thermodynamics states that {\em the entropy of a closed system never decreases}. If entropy is denoted by $S$, this law reads:
\begin{equation}
	\frac{dS}{dt}\geq 0.
\end{equation}

In the 1870s, Ludwig Boltzmann (1844-1906) argued that the effect of randomly moving gas molecules was to ensure that the entropy of a gas would increase, until it reached its maximum possible value. This is his famous {\em H-theorem}. Boltzmann was able to show that macroscopic distributions of great inhomogeneity (i.e. of high order or low entropy) are formed from relatively few microstate arrangements of molecules, and were, consequently, relatively improbable. Since physical systems do not tend to go into states that are less probable than the states that they are in, it follows that any system would evolve toward the macrostate that is consistent with the larger number of microstates. The number of microstates and the entropy of the system are related by the fundamental formula:
\begin{equation}
	S= k \ln W,
\end{equation}
where $k=10^{-23}$ JK$^{-1}$ is Boltzmann's constant and $W$ is the volume of the phase-space that corresponds to the macrostate of entropy $S$. 

More than twenty years after the publication of Boltzmann's fundamental papers on kinetic theory, it was pointed out by Burbury \cite{Burbury1},\cite{Burbury2} that the source of asymmetry in the H-theorem is the assumption that the motions of the gas molecules are independent before they collide and not afterward, if entropy is going to increase. This essentially means that the entropy increase is a consequence of the {\em initial conditions} imposed upon the state of the system. Boltzmann's response was: 

\begin{quotation}
There must then be in the Universe, 
which is in thermal equilibrium as a 
whole and therefore dead, here and 
there, relatively small regions of the 
size of our  world, which during the 
relatively short time of eons deviate 
significantly from thermal equilibrium.  
Among these worlds the state probability 
increases as often as it decreases \cite{Boltzmann}.        
\end{quotation} 

As noted by Price \cite{Price}: ``The low-entropy condition of our region seems to be associated entirely with a low-energy condition in our past.''

The probability of the large fluctuations required for the formation of the Universe we see, on the other hand, seems to be zero, as noted long ago by Eddington (1931): ``A Universe containing mathematical physicists 
 at any assigned date will be in the state of 
maximum disorganization which is not inconsistent 
with the existence of such creatures.'' Large fluctuations are rare ($P\sim \exp{-\Delta S}$); {\em extremely} large fluctuations, basically impossible . For the whole Universe, $\Delta S\sim 10^{104}$ in units of $k=1$. This yields $P=0$.

In 1876, a former teacher of Boltzmann and later colleague at the University of Vienna, J. Loschmidt, noted:

\begin{quotation}
Obviously, in every arbitrary system the course of events must become retrograde when the velocities of all its elements are reversed \cite{Loschmidt}.
\end{quotation}

Putting the point in modern terminology, the laws of (Hamiltonian) mechanics are such that for every solution one can construct another solution by reversing all velocities and replacing $t$ by $-t$. Since the Boltzmann's function $H[f]$ is invariant under velocity reversal, it follows that if $H[f]$ decreases for the first solution, it will increase for the second. Accordingly, the reversibility objection is that the H-theorem cannot be a general theorem for all mechanical evolutions of the gas. More generally, the problem goes far beyond classical mechanics and encompasses our whole representation of the physical world. This is because {\em all formal representations of all fundamental laws of physics are invariant under the operation of time reversal}. Nonetheless, the evolution of all physical processes in the Universe is irreversible. 

If we accept, as mentioned, that the origin of the irreversibility is not in the laws but in the initial conditions of the laws, two additional problems emerge: 1) What were exactly these initial conditions?, and 2) How the initial conditions, of global nature, can enforce, at any time and any place, the observed local irreversibility? 

The first problem is, in turn, related to the following one, once the cosmological setting is taken into account: in the past, the Universe was hotter and at some point matter and radiation were in thermal equilibrium; how is this compatible with the fact that entropy has ever been increasing according to the so-called Past Hypothesis, i.e. entropy was at a minimum at some past time and has been increasing ever since?  

The standard answer to this question invokes the expansion of the Universe: as the Universe expanded, the maximum possible entropy increased with the size of the Universe, but the actual entropy was left well behind the permitted maximum. The Second Law of Thermodynamics and the source of irreversibility is the trend of the entropy to reach the permitted maximum. According to this view, the Universe actually began in a state of maximum entropy, but due to the expansion, it was still possible for the entropy to continue growing.        

The main problem with this line of thought is that it is not true that the Universe was in a state of maximum disorder at some early time. In fact, although locally matter and radiation might have been in thermal equilibrium, this situation occurred in a regime where the global effects of gravity cannot be ignored \cite{Penrose}. Since gravity is an attractive force, and the Universe was extremely smooth (i.e structureless) in early times, as indicated, for instance, by the measurements of the cosmic microwave background radiation, the gravitational field should have been quite far from equilibrium, with very low global entropy \cite{Penrose}. It seems, then, that the early Universe was {\em globally} out of the equilibrium, being the total entropy dominated by the entropy of the gravitational field. If we denote by $C^{2}$ a scalar formed out by contractions of the Weyl tensor, the initial condition $C^{2}\sim 0$ is required if entropy is still growing today \footnote{This is because the Weyl tensor provides a measure of the inhomogeneity of the gravitational field.}.  

How the Second Law is locally enforced  by the initial conditions, which are of a global nature? A possible answer is that there should be a coupling between gravitation (of global nature) and electrodynamics (of local action). 

The electromagnetic radiation field can be described in the terms of the 4-potential $A^{\mu}$, which in the Lorentz gauge satisfies:
\begin{equation}
\partial^{\nu}\partial_{\nu}A^{\mu}(\vec{r},\;t)=4\pi j^{\mu} (\vec{r},\;t),
\end{equation}
with $c=1$ and $j^{\mu}$ the 4-current. The solution $A^{\mu}$ is a functional of the sources $j^{\mu}$. The retarded and advanced solutions are:

\begin{equation}
	A^{\mu}_{\rm ret}(\vec{r},\;t)=\int_{V_{\rm ret}}
\frac{j^{\mu} \left(\vec{r},\;t-\left|\vec{r}-\vec{r'}\right|\right)}{\left|\vec{r}-\vec{r'}\right|}d^{3}\vec{r'} + \int_{\partial V_{\rm ret}}
\frac{j^{\mu} \left(\vec{r},\;t-\left|\vec{r}-\vec{r'}\right|\right)}{\left|\vec{r}-\vec{r'}\right|}d^{3}\vec{r'}, \label{ret}
\end{equation}

\begin{equation}
	A^{\mu}_{\rm adv}(\vec{r},\;t)=\int_{V_{\rm adv}}
\frac{j^{\mu} \left(\vec{r},\;t+\left|\vec{r}-\vec{r'}\right|\right)}{\left|\vec{r}-\vec{r'}\right|}d^{3}\vec{r'} + \int_{\partial V_{\rm adv}}
\frac{j^{\mu} \left(\vec{r},\;t+\left|\vec{r}-\vec{r'}\right|\right)}{\left|\vec{r}-\vec{r'}\right|}d^{3}\vec{r'}. \label{adv}
\end{equation}

The two functionals of $j^{\mu}(\vec{r},\;t)$ are related to one another by a time reversal transformation. The solution (\ref{ret}) is contributed by sources in the past of the space-time point $p(\vec{r},\;t)$ and the solution (\ref{adv}) by  sources in the future of that point. The integrals in the second term on the right side are the surface integrals that give the contributions from i) sources outside of $V$ and ii) source-free radiation. If $V$ is the causal past and future, the surface integrals do not contribute. At this point it is convenient to introduce formal definitions of causal curves, past and future \cite{H-E}.\\

{\bf Definition.} {\sl A causal curve in a space-time $(M,\; g_{\mu\nu})$ is a curve that is non space-like, that is, piecewise either time-like or null (light-like).} \\

{\bf Definition.} {\sl If $(M,\; g_{\mu\nu})$ is a time-orientable space-time, then $\forall p\in M$, the causal future of $p$, denoted $J^{+}(p)$, is defined by:

\begin{equation}
	J^{+}(p)\equiv \left\{ q \in M | \exists \;a\; future-directed \; causal \; curve \; from  \; p \; to \; q \right\}.
\end{equation}
} 

Similarly,\\

{\bf Definition.} {\sl If $(M,\; g_{\mu\nu})$ is a time-orientable space-time, then $\forall p\in M$, the causal past of $p$, denoted $J^{-}(p)$, is defined by:

\begin{equation}
	J^{-}(p)\equiv \left\{ q \in M | \exists \;a\; past-directed \; causal \; curve \; from  \; p \; to \; q \right\}.
\end{equation}
} \\
  
We see, then, that if $V_{\rm adv}=J^{+}(p)$ and $V_{\rm ret}=J^{-}(p)$ there are no contributions from outside of the causal past and future, and the surface integrals yield zero.
  
The linear combinations of electromagnetic solutions are also solutions, since the equations are linear and the Principle of Superposition holds. It is usual to consider only the retarded potential as physically meaningful in order to estimate the electromagnetic field at $p(\vec{r},\;t)$: $F^{\mu\nu}_{\rm ret}=\partial^{\mu}A^{\nu}_{\rm ret}-\partial^{\nu}A^{\mu}_{\rm ret}$. However, there seems to be no compelling reason for such a choice. We can adopt, for instance (in what follows we use a simplified notation),
\begin{equation}
	A^{\mu}(\vec{r},\;t)=\frac{1}{2}\left(\int_{J^{+}} {\rm adv\;} + \;\int_{J^{-}} {\rm ret}\right)\; dV.
\end{equation}
 
If the space-time is curved ($R\neq 0$), the null cones that determine the local causal structure will not be symmetric around the point $p$ $(\vec{r},\;t)$. In particular, the presence of cosmological particle horizons can make the contributions very different from both integrals. Particle horizons occur whenever a particular system never gets to be influenced by the whole space-time.\\

{\bf Definition.} {\sl For a causal curve $\gamma$ the associated future (past) particle horizon is defined as the boundary of the region from which the causal curves can reach some point on $\gamma$.} \\

Finding the particle horizon (if one exists at all) requires a knowledge of the global space-time geometry. Particle horizons occur in systems undergoing lasting acceleration. The radius of the past particle horizon  is \cite{Rindler}:

\begin{equation}
	R_{\rm past}= a(t) \int^{t}_{t'=0} \frac{c}{a(t')} dt',
\end{equation}
where $a(t)$ is the time-dependent scale factor of the Universe. The radius of the future particle horizon (some times called event horizon) is:

\begin{equation}
	R_{\rm future}= a(t_0) \int^{\infty}_{t'_0} \frac{c}{a(t')} dt'.
\end{equation}

If the Universe is accelerating, as seems to be suggested by recent observations \cite{Perlmutter}, then $J^{+}(p)$ and $J^{-}(p)$ are not symmetric and the result of integrals (\ref{ret}) and (\ref{adv}) is different.   
We can then introduce a vector field $L^{\mu}$ given by:  
   
\begin{equation}
L^{\mu}= \left[\int_{J^{-}} {\rm ret} - \int_{J^{+}} {\rm adv} \right]\; dV \neq 0.
\end{equation}

If $g_{\mu\nu}L^{\mu}T^{\nu}\neq0$, with $T^{\nu}=(1,0,0,0)$ there is a preferred direction for the Poynting flux in space-time. The Poynting flux is given by:

\begin{equation}
	\vec{S}=\frac{4\pi}{c} \vec{E} \times \vec{B}= (T^{01}_{\rm EM}, T^{02}_{\rm EM}, T^{03}_{\rm EM}),
\end{equation}
where $\vec{E}$ and $\vec{B}$ are the electric and magnetic fields and $T^{\mu\nu}_{\rm EM}$ is the electromagnetic energy-momentum tensor.

In a black hole interior the direction of the Poynting flux is toward the singularity. In an expanding, accelerating Universe, it is in the global future direction. We see, then, that a time-like vector field, in a general space-time $(M, g_{\mu\nu})$, can be {\sl anisotropic}. There is a global to local relation given by the Poynting flux as determined by the curvature of space-time that indicates the direction along which events occur. Physical processes, inside a black hole, have a different orientation from outside, and the causal structure of the world is determined by the dynamics of space-time and the initial conditions. Macroscopic irreversibility\footnote{Notice that the electromagnetic flux is related with the macroscopic concept of temperature through the Stefan-Boltzmann law: $L=A\sigma_{\rm SB}T^{4}$, where $\sigma_{\rm SB}= 5.670 400 \times 10^{-8} \textrm{J\,s}^{-1}\textrm{m}^{-2}\textrm{K}^{-4}$ is the 
Stefan-Boltzmann constant.} and time anisotropy emerge from fundamental reversible laws. 

There is an important corollary to these conclusions. Local observations about the direction of events can provide information about global features of space-time and the existence of horizons and singularities.

\section{Relativity and non-locality}

The compatibility of Relativity theory and quantum mechanics is an essential issue in the foundations of physical science. The experimental refutation of Bell's inequalities\footnote{Bell's theorem states that {\em no physical theory which is realistic and also local in a specified sense can agree with all of the statistical implications of Quantum Mechanics}. Here the word `realistic' actually means deterministic in a Laplacian sense (see above). Many different versions and cases, with family resemblances, were inspired by Bell's 1964 paper \cite{Bell} and are subsumed under the italicized statement, ``Bell's Theorem'' being the collective name for the entire family.} in the 1980s indicated that quantum mechanics is not a local theory since action-at-distance effects are present. This seems to contradict one of the basic postulates of Special (and General) Relativity: the impossibility for any physical system to surpass the speed of light. In this section we will take a closer look at this problem.   

\subsection{The problem: quantum entanglement}

Quantum mechanics embraces action at a distance with a property called entanglement, in which two particles behave synchronously with no intermediary; it is a nonlocal property. For instance, calcium vapor exposed to lasers fluoresces. Excited atoms cascade down to their ground states and they give off light. Each atom emits a pair of photons which travel off in opposite directions. The polarization of these photons shows no preferred direction. The pairs, however, display a striking correlation: each member of the pair always acts as if it has the same polarization as its partner. This quantum connection has different properties:

\begin{itemize}
\item The quantum connection is unattenuated by distance.
\item The quantum connection is discriminating: only particles which have interacted in the past are affected by it. No classical force exhibits this behavior. 
\item The quantum connection is faster than light, and likely instantaneous. This seems to be incompatible with relativistic space-time structure. 
\end{itemize}

Everybody knows that Relativity forbids something. Not everyone agrees, however, about what is forbidden. There are different possibilities discussed in the literature, as reviewed by Tim Maudlin \cite{Maudlin}:

\begin{enumerate}
\item Relativity forbids matter to be transported faster than light.
\item Relativity forbids signals to be sent faster than light.
\item Relativity forbids information to be transmitted faster than light.
\item Relativity forbids causal processes to propagate faster than light.
\end{enumerate} 

Actually, Relativity just states that subluminal systems cannot become superluminal and viceversa. Hypothetical superluminal tachyons do not violate Relativity's laws. Tachyon theory shows that Relativity {\em is not} restricted to systems with subluminal speed in order to be consistent. On other hand, violation of Bell's inequalities does not require superluminal matter transport or signaling. Much less of information.
Information is a property of languages, not of physical systems.  Contrary to what Tim Maudlin states \cite{Maudlin}, superluminal transmission of information is not required by quantum entanglement. Information has not, and cannot, have any effect upon physical systems.

Contrary to all these views, I sustain that violation of Bell's inequalities is possible if causation can be non-local. In what follows I provide a non-local definition of causation (see Ref. \cite{Romero-Perez} for more details).

\subsection{Causality revisited}

Causation is a form of event generation \cite{Bunge1979}. To present an explicit definition of causation requires introducing some ontological concepts to formally characterize what is understood by `event'.

The concept of individual is the basic primitive concept of any ontological theory\footnote{I follow Bunge \cite{Bunge1977} on the basics of the ontological views presented here.}. Individuals associate themselves with other individuals to yield new individuals. It follows that they satisfy a calculus, and that they are rigorously characterized only through the laws of such a calculus. These laws are set with the aim of reproducing the way real things associate. Specifically, it is postulated that every individual is an element of a set $s$ in such a way that the structure $\textsl{S}=\left\langle s, \circ, \square \right\rangle$ is a \textit{commutative monoid of idempotents}. This is a simple additive semi-group with neutral element.

In the structure \textsl{S}, $s$ is the set of all individuals, the element $\square \in s$ is a fiction called the null individual, and the binary operation $\circ$ is the association of individuals. Although \textsl{S} is a mathematical entity, the elements of $s$ are not, with the only exception of $\square$, which is a fiction introduced to form a calculus. The association of any element of $s$ with $\square$ yields the same element. The following definitions characterize the composition of individuals.

\begin{enumerate}

\item ${x} \in s $ is composed $ \Leftrightarrow  \left(\exists {y}, {z}\right)_{s} \left( {x} ={y} \circ {z} \right) $

\item ${x} \in s $ is simple $ \Leftrightarrow \; \sim \left(\exists {y}, {z}\right)_{s} \left({x} ={y} \circ {z} \right)$

\item $ {x}\subset {y}\ \Leftrightarrow {x} \circ {y} = {y}\ $ (${x}$ is part of ${y}\ \Leftrightarrow {x} \circ {y} = {y} $) 

\item $ \textsl{Comp}({x}) \equiv\{{y}\in s \;|\; {y}\subset {x}\}$ is the composition of ${x}$.\\ 

\end{enumerate}

Real things are distinguished from abstract individuals because they have a number of properties in addition to their capability of association. These properties can be \textit{intrinsic} \normalfont ($P_i$) or \textit{relational} \normalfont ($P_r$). The intrinsic properties are inherent and they are represented by predicates or unary applications, whereas relational properties depend upon more than a single thing and are represented by $n$-ary predicates, with $n \geq 1$. Examples of intrinsic properties are electric charge and rest mass, whereas velocity of macroscopic bodies and volume are relational properties\footnote{Velocity is an intrinsic property only in the case of photons and other bosons that move at the speed of light in any reference system.}.

An individual with its properties make up a thing $X$:
\[
	X=<x,P(x)>
\]	
	
Here $P(x)$ is the collection of properties of the individual $x$. A material thing is an individual with concrete properties, {\em i.e.} properties that can change (see below) in some respect.

The {\em state} of a thing $X$ is a set of functions $S(X)$ from a 
domain of reference $M$ (a set that can be enumerable or nondenumerable\footnote{In most physically interesting cases $M$ is a space-time continuum.}) to the set of properties ${\cal P}_{X}$. Every function in $S(X)$ represents a property in ${\cal P}_{X}$. The set
of the {\sl physically accessible} states of a thing $X$ is the {\em lawful state space} of
$X$: $S_{\rm L}(X)$. The state of a thing is represented by a point in
$S_{\rm L}(X)$. A change of a thing is an ordered pair of states. Only changing things can be material. Abstract things cannot change since they have only one state (their properties are fixed by definition).

A {\em legal statement} is a restriction  upon the state
functions of a given class of things. A {\em natural law}
is a property of a class of material things represented by an empirically corroborated legal statement.

The {\em ontological history} $h(X)$ of a thing $X$ is a subset of
$S_{\rm L}(X)$ defined by
\[
	h(X) = \{ \langle t, F(t) \rangle | t \in M\} 
\] 
where $t$ is an	element of some auxiliary set $M$,
and $F$ are the functions that represent the properties 
of $X$.

If a thing is affected by other things we can introduce the following definition:\\

$h(Y/X)$: ``history of the thing $Y$ in presence of the thing $X$''.\\

Let $h(X)$ and $h(Y)$ be the histories of the things $X$ and $Y$, respectively.\\
Then 
\[
	h(Y/X) = \{ \langle t, H(t) \rangle |\: t \in M\}, 
\] 
where $ H\neq F $ is the total state function of $Y$ as affected by the existence of $X$, and $F$ is the total state function of $X$ in the absence of $Y$. The history of $Y$ in presence of $X$ is different from the history of $Y$ without $X$.\\

We can now introduce the notion of {\sl action}:\\

$ X \triangleright Y $: ``$X$ acts on $Y$''
\[
 X \triangleright Y \stackrel{def}{=} h(Y/X)\neq h(Y) 
\] 

 An {\sl event} is a change of a thing $X$, {\em i.e.} an ordered pair of states:
\[ (s_1, s_2 ) \in E_{\rm L}(X) = S_{\rm L}(X) \times S_{\rm L}(X) \]

The space $E_{\rm L}(X)$ is called
the {\em event space} of $X$.

\textit{Causality} is a relation between events, {\em i.e.} a relation between changes of states of concrete things. It is {\sl not} a relation between things. Only the related concept of `action' is a relation between things. Specifically,\\

$ \mathfrak{C}(x)$: ``an event in a thing $x$ is caused by some unspecified event $e^{x}_{x_{i}}$''.

$$
\mathfrak{C}(x)\stackrel{def}{=} (\exists e^{x}_{x_{i}}) \left[ e^{x}_{x_{i}}\in E_{\rm L}(x)\right] \Leftrightarrow x_{i} \triangleright x.
$$

$ {C}(x, y)$: ``an event in a thing $x$ is caused by an event in a thing $y$''.

$$
{C}(x, y)\stackrel{def}{=} (\exists e^{x}_{y}) \left[ e^{x}_{y}\in E_{\rm L}(x) \right] \Leftrightarrow y \triangleright x.
$$

In the above definitions\footnote{These definitions are part of joint work with Daniela P\'erez.}, the notation $e^{x}_{y}$ indicates in the superscript the thing $x$ to whose event space belongs the event $e$, whereas the subscript denotes the thing that acted triggering the event. The implicit arguments of both $\mathfrak{C}$ and $C$ are events, not things. For simplicity in the notation we refer to the things that undergo the events. We shall also use hereafter lower case letters for variables that take values upon a domain of things.  

Causation is a form of event generation. The crucial point is that a given event in the lawful event space $E_{\rm L}(x)$ is caused by an action of a thing $y$ iff the event happens only conditionally to the action, {\em i.e.}, it would not be the case of $e^{x}_{y}$ without an action of $y$ upon $x$. Time does not appear in this definition, allowing causal relations in space-time without a global time orientability or even instantaneous and non-local causation.

If causation is non-local under some circumstances, e.g. when a quantum system is prepared in a specific state of polarization or spin, quantum entanglement poses no problem to realism and determinism. The quantum theory describes an aspect of a reality that is ontologically determined and with non-local relations. Under any circumstances the postulates of Special Relativity are violated, since no physical system ever crosses the barrier of the speed of light.   

\section{Further issues}

In this article it was only possible to make a quick survey of some issues in the foundations of space-time theories. Many more can be mentioned. For instance,
\begin{itemize} 
\item {\em What is the ontological nature of singularities in physical theories?} Singularities appear in space-time theories but they do not belong to the reference domain of the theories, since the manifold model of space-time breaks if it does not remain regular. What is the physics hidden by the singularities? 
\item {\em Paradoxes generated by close timelike curves (CTCs)}. These paradoxes (e.g. see Ref. \cite{Romero2010}) seem to suggest the existence of consistency constraints at an ontological level. 
\item {\em Zeno-like paradoxes}. These paradoxes are related to the notions of continuum and metric properties of sets. 
\item {\em Identity}. How is possible to say that my dog is the same dog that my daughter bought as a puppy two years ago? I do not remember anything about myself when I was 1 year old. In what sense I am the same person? What if I lose all my memories? Shall I still be the same person? What is a person? What is identity through time?  
\item {\em Constitution}: When change of constitution changes identity?
\item {\em Spacetime and God}. Is the concept of God compatible with the manifold model of space-time? If the world is determined, God knows all events. How can He change something then? What happens with His omnipotence if He cannot change anything?  
\item {\em The nature of explanation}. How complete are our explanations of the events? How much can we know?
\item {\em Representation and reality}. Can we know the true ontology of the world? Can an ontology be {\em true}?
\item {\em Ethics}: How we should live in a world like this?
\end{itemize}

Just a few questions, of many more that can be posed. I dare suggest, nonetheless, that none is as much important for us as the latter. For the Hellenistic Greeks, physics was a road to a better ethics. For most of us, regrettably, it is not even a road. It is just a job.    

\section{Conclusions}

We have looked at some issues in the foundations of space-time theories. The main conclusions can be clearly stated. 
There seem to be global-to-local connections that generate what we call `irreversibility', which is essentially asymmetry in space-time. 
Causation seems to be a non-local event generation process for things that form a system (i.e., connected things). 
There is no `becoming'. The world is just the totality of all events of all things. 
Philosophical inquiry on the foundations of scientific theories is essential to determine the consistency, scope, reference, and assumptions of our representations of the physical world. Better a conjectural, explicit philosophy, that can be criticized, than a bunch of hidden prejudices that can only be defended.

%%%%%%%%%%%%%%%%%%%%%%%%%%%%%%%%%%%%%%%%%%%%%%%%
%% BACKMATTER
%%%%%%%%%%%%%%%%%%%%%%%%%%%%%%%%%%%%%%%%%%%%%%%%

\begin{theacknowledgments}
I am very grateful to Professor Mario Novello for his kind invitation to participate in this stimulating school. I thank Professors Mario Bunge, J.M. Salim,  Nelson Pinto Neto, Felipe Tovar Falciano, Marc Lachieze-Rey, and H. Vucetich, as well as my student Daniela P\'erez  for constructive criticisms and comments. I am deeply indebted to my dear friend Professor Santiago E. Perez Bergliaffa for 25 years of discussions on philosophy, science, and life. The road has not been so lonely thanks to him.  I was supported by research grant PIP 0078 from CONICET.

\end{theacknowledgments}



\bibliographystyle{aipproc}   % if natbib is available

\begin{thebibliography}{9}

\bibitem{Bunge1967}
M. ~ Bunge, \emph{Foundations of Physics}, Springer-Verlag, Berlin, 1967.

\bibitem{Borges}
J.~L. Borges, \emph{Historia de la Eternidad}, Emec\'e, Buenos Aires, 1953.

\bibitem{Lindberg}
D.~C. Lindberg, \emph{The Beginnings of Western Science: The European Scientific Tradition in Philosophical, Religious, and Institutional Context, Prehistory to A.D. 1450 }, University Of Chicago Press, 2 edition, Chicago,  2008. 

\bibitem{Ariew}
Ariew, R., (ed.), \emph{ G. W. Leibniz and Samuel Clarke. Correspondence},  Hackett, Indianapolis, 2000.

\bibitem{Mach}
E. Mach, \emph{The Science of Mechanics;  a Critical and Historical Account of its Development }, LaSalle, Ill., Open Court Pub. Co., 1960.

\bibitem{Einstein1}
A. Einstein,  \emph{Annalen der Physik} \textbf{322},  891--321  (1905).

\bibitem{Minkowski}
H. Minkowski,  Lecture ``Raum und Zeit, 80th Versammlung Deutscher Naturforscher (K\"oln, 1908)",  \emph{Physikalische Zeitschrift} {\bf 10}, 75-88  (1909).

 \bibitem{Einstein2}
A. Einstein, \emph{Annalen der Physik} {\bf 49}, 769-822 (1916).

\bibitem{Rovelli}
C. ~ Rovelli, \emph{Quantum Gravity}, Cambridge University Press, Cambridge, 2004.

\bibitem{Bergliaffa}
S.E.  Perez Bergliaffa, G. E. Romero,  and H. Vucetich,  \emph{  Int. J. Theor. Phys.} \textbf{37},  2281--2298  (1998).

\bibitem{Reichenbach}
H. Reichenbach, \emph{The Philosophy of Space and Time }, Dover, New York,  1958.


\bibitem{Grunbaum}
A.  Gr\"unbaum, \emph{Philosophical Problems of Space and Time}, Reidel, 2nd ed., Dordrecht, 1973.


\bibitem{Smart}
J.J.C. Smart, \emph{Philosophy and Scientific Realism}, Routledge and Kegan Paul, New York, 1963.

\bibitem{Weyl}
H. Weyl, \emph{Philosophy of Mathematics and Natural Science}, Princeton University Press, Princeton, 1949.

\bibitem{Earman}
J. Earman, \emph{A Primer on Determinism}, Reidel, Dordrecht, 1986.

\bibitem{Burbury1}
S.H. Burbury, \emph{Nature} {\bf 51}, 78-79 (1894).

\bibitem{Burbury2}
S.H. Burbury, \emph{Nature} {\bf 51}, 320-320 (1895).

\bibitem{Boltzmann}
L. Boltzmann, \emph{Nature} {\bf 51}, 413-415 (1895).

\bibitem{Price} H. Price, in: \emph{Contemporary Debates in Philosophy of Science}, C. Hitchcock (ed.), Blackwell, Singapore, p. 21, 2004.

\bibitem{Eddington}
A.S. Eddington, \emph{Nature} {\bf 127}, 447-453 (1931).

\bibitem{Loschmidt} 
J. Loschmidt, \emph{Wiener Berichte} {\bf 73}, 128-142 (1876).

\bibitem{Penrose} 
R. Penrose, in \emph{General Relativity: An Einstein Centennial}, S.W. Hawking \& W. Israel (eds.), Cambridge University Press, Cambridge, p.581, 1979. 

\bibitem{H-E} 
S.W. Hawking, and G.F.R. Ellis, \emph{The Large Scale Structure of Space-Time}, Cambridge University Press, Cambridge, 1973. 


\bibitem{Rindler}
W. Rindler, \emph{Monthly Notices of the Royal Astronomical Society} {\bf 116}, 662-667 (1956).

\bibitem{Perlmutter} 
S. Perlmutter, et al.,  \emph{The Astrophysical Journal} {\bf 517}, 565-586 (1999).

\bibitem{Maudlin}
T. Maudlin, \emph{Quantum Nonlocality and Relativity}, Blackwell, Oxford, 1994.

\bibitem{Bell}
J.S. Bell, \emph{Physics} {\bf 1}, 195-200 (1964).

\bibitem{Romero-Perez}
G.E. Romero and D. P\'erez, \emph{Bolet{\'{i}}n de la Asociaci\'on Argentina de Astronom{\'{i}}a } {\bf 52}, 221-224 (2009).

\bibitem{Bunge1979}
M. ~ Bunge, \emph{Causality and Modern Science}, Dover, New York, 1979.

\bibitem{Bunge1977}
M. ~ Bunge, \emph{Ontology I: The Furniture of the World}, Kluwer, Dordrecht, 1977.

\bibitem{Romero2010}
G.E. ~ Romero, \emph{Es Posible Viajar en el Tiempo?}, Kaicron, Buenos Aires, 2010.


\end{thebibliography}


\end{document}